\documentclass[%
 reprint,
 superscriptaddress,
onecolumn,
nofootinbib,
 amsmath,amssymb,
 aps,
floatfix,
]{revtex4-2}

\usepackage{graphicx}
\usepackage{dcolumn}
\usepackage{bm}
\usepackage{xcolor}
\usepackage{hyperref}
\usepackage{placeins}
\usepackage[mathlines]{lineno}
\usepackage{setspace}
\usepackage[version=4]{mhchem}
\usepackage{multirow}

\newcommand{\dd}{\mathrm{d}}

\begin{document}

\title{Nonparametric multiscale modeling of boundary lubrication: hexadecane in highly pressurized gold asperity contacts}

\author{Hannes Holey}
\affiliation{Department of Physics, Center for Complexity and Biosystems, University of Milan, Via Celoria 16, 20133 Milan, Italy}

\author{Michael Moseler}
\affiliation{Fraunhofer IWM, MicroTribology Centrum µTC, Wöhlerstraße 11, 79108 Freiburg, Germany}
\affiliation{Freiburg Materials Research Center, University of Freiburg, Stefan-Meier-Straße 21, 79104 Freiburg, Germany}
\affiliation{Institute of Physics, University of Freiburg, Hermann-Herder-Straße 3a, 79104 Freiburg, Germany}

\author{Peter Gumbsch}
\affiliation{Fraunhofer IWM, MicroTribology Centrum µTC, Wöhlerstraße 11, 79108 Freiburg, Germany}
\affiliation{Institute for Applied Materials, Karlsruhe Institute of Technology, Straße am Forum 7, 76131 Karlsruhe, Germany}

\author{Lars Pastewka}
\email{lars.pastewka@imtek.uni-freiburg.de}
\affiliation{Department of Microsystems Engineering (IMTEK), University of Freiburg, Georges-Köhler-Allee 103, 79110 Freiburg, Germany}

\date{\today}

\begin{abstract}
Boundary lubrication is governed by molecular processes at the sliding interface that are inaccessible to classical continuum descriptions.
While molecular dynamics (MD) simulations can resolve these processes in atomistic detail, incorporating their output into engineering-scale models through semi-empirical constitutive laws becomes increasingly difficult as confinement approaches the molecular scale. 
Here, we apply a nonparametric multiscale framework based on Gaussian process (GP) regression to model boundary lubrication of hexadecane confined between gold surfaces under pressures up to 1 GPa and gap heights down to 1.4 nm. 
The GP surrogates are trained on nonequilibrium MD simulations of the confined fluid and directly provide stress predictions to a continuum thin-film solver, circumventing the need for fixed-form constitutive laws.
The framework naturally captures molecular phenomena such as density layering, viscosity changes, and the strongly nonlinear wall slip that dominates the frictional response at high pressures and small gap heights.
Our results reproduce the atomistic benchmark data of \citet{codrignani2023_continuum}, demonstrating that nonparametric surrogate models offer a flexible and physically transparent route toward predictive continuum modeling of boundary lubrication.
\end{abstract}

                              
\maketitle

\section{Introduction}

Boundary lubrication is becoming increasingly relevant in modern tribological systems, driven by the push toward low-viscosity lubricants under environmental constraints \cite{meng2020_review}, and by a shift in lubricant performance requirements demanded by high load-bearing applications such as electric vehicle drivetrains \cite{farfan-cabrera2019_tribology} or wind turbine gearboxes \cite{kotzalas2010_tribological}.
At the engineering scale, the transition to mixed and boundary lubrication is often characterized by a lubricant film thickness on the order of the (deformed) asperity heights of the rough surfaces \cite{hansen2021_new}.
Solvers for elastohydrodynamic lubrication (EHL) have been extended to these regimes by assuming dry friction in regions where the fluid film thickness falls below a threshold, typically on the order of tens to hundreds of nanometers \cite{hu2000_full,jiang1999_mixeda}.
Thus, these models neglect the contribution to overall friction from lubricant films in the nanometer range and below \cite{archard1962_lubrication,glovnea2003_measurement}, where fluid properties deviate from those of the bulk, since structural changes \cite{horn1981_direct,gee1990_liquid} and wall slip \cite{pit2000_direct,zhu2001_ratedependent,cheng2002_fluid} become important.

As the film thickness approaches the size of the lubricant molecules, molecular specificity can no longer be neglected.
Molecular dynamics (MD) simulations have been widely used in tribology to address this issue \cite{ewen2018_advances}.
For instance, MD simulations of bulk fluids enable the characterization of lubricant rheology at high pressure and shear rates \cite{jadhao2019_rheological}, i.e., conditions similar to those found in EHL contacts.
Furthermore, simulations of confined fluid systems provide insight into flow boundary conditions, where density layering \cite{thompson1990_shear,gao1997_layering} and wall slip \cite{thompson1997_general,savio2016_boundary} prevail.
Incorporating these insights into continuum models requires choosing a constitutive law that describes how bulk or surface properties change with local operating conditions \cite{martini2006_molecular,savio2015_multiscale, codrignani2023_continuum}.
The choice of the functional form represents a trade-off between empirical accuracy and physical interpretability \cite{gao2023_shearthinning}, and has been extensively debated in the EHL community \cite{spikes2014_historya,bair2015_comment,spikes2015_reply}.

Interatomic potentials used in MD simulations have become increasingly accurate, enabling more realistic descriptions of lubricants and interfaces at the molecular scale \cite{ewen2021_contributions}.
For the incorporation of molecular behavior into predictive multiscale models, the specific functional form or constitutive relation by which this behavior is described is of secondary importance, provided it accurately reproduces the underlying data.
This shifts the focus toward capturing combined effects, such as shear thinning, shear heating, and wall slip as competing mechanisms at a tribological interface \cite{peeters2026_when}, rather than considering them in isolation.
Machine learning offers an alternative to fixed-form constitutive models, where one can train a surrogate model directly on MD data and let the data determine the appropriate level of complexity.
Gaussian process (GP) regression \cite{rasmussen2006_gaussian} is particularly well suited to this role, as it provides not only flexible interpolation between MD observations but also a built-in measure of prediction uncertainty, which can be exploited to guide the acquisition of new training data.

In our previous work \cite{holey2025_active}, we developed a concurrent multiscale framework in which GP surrogates replace fixed-form constitutive laws entirely, and an active learning algorithm ensures that MD simulations are run only where the surrogate is insufficiently accurate.
The surrogate models are trained directly on shear and normal stress measured as wall traction in nonequilibrium MD simulations, circumventing the need for intermediate variables such as viscosity or slip length, which lose their unambiguous definition when the gap contains only a few molecular layers.
Here, we apply this multiscale framework to the hexadecane-gold nanoasperity contact studied by \citet{codrignani2023_continuum}, for which full-scale atomistic benchmark data in a wide range of  applied shear rates and normal loads exist.
We use active learning to train the surrogate models on nonequilibrium confined MD systems that account for the heterogeneous surface properties found in the system.
Then, we compare the obtained pressure and shear stress profiles with the full-scale reference and with Reynolds calculations based on the parametric constitutive laws from Ref.~\cite{codrignani2023_continuum}.
Finally, we compare the nonparametric surrogate models with the parametric ones, and test the transferability of the obtained models to systems outside the training set.

\section{Methods}

\begin{figure*}[!ht]
    \centering
    \includegraphics[width=\linewidth]{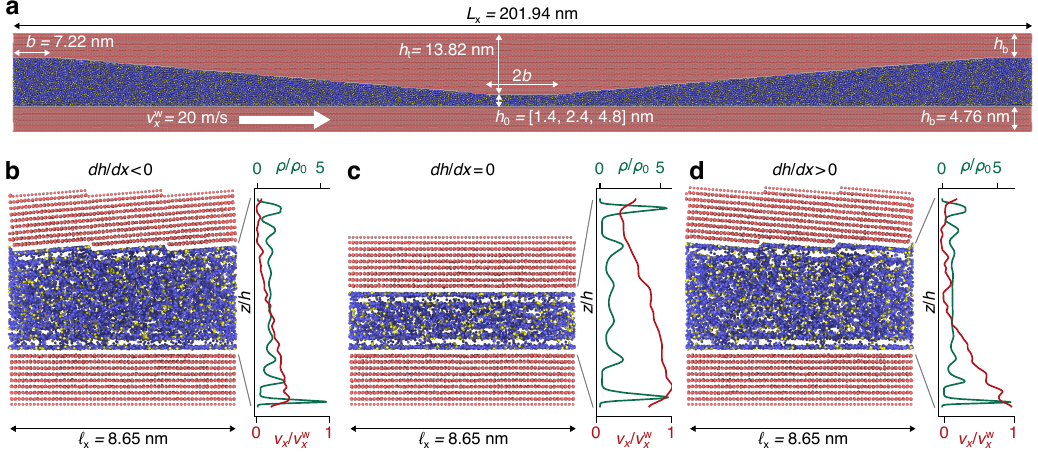}
    \caption{%
    \textbf{Geometry of the converging diverging channel (CDC) and the representative volume elements used for the training simulations.}
    \textbf{a} Setup of the full-scale atomistic CDC simulations from \citet{codrignani2023_continuum}. No additional full-scale simulations have been performed in this work.
    \textbf{b} MD setup of the representative volume element for the converging section of the CDC with heterogeneous surface properties. The step-patterned top wall has been generated by a simple shear deformation corresponding to the inclination angle of the gap profile. The red line shows the velocity profile across the channel height, and the green line illustrates the density profile.
    \textbf{c} Same as \textbf{b} but for homogeneous, ideally flat (111) surfaces corresponding to the center or side regions of the CDC.
    \textbf{d} Same as \textbf{b} but for the diverging section of the CDC.
    }
    \label{fig:cdc}
\end{figure*}

We model lubricated fluid flow through the converging-diverging channel (CDC) geometry introduced in Ref.~\cite{codrignani2023_continuum}.
Figure~\ref{fig:cdc}a shows the channel geometry, which can be seen as an idealized, lubricated asperity contact of rough surfaces typical for mixed and boundary lubrication with minimum gap heights of only a few nanometers.
\citet{codrignani2023_continuum} performed large scale molecular dynamics (MD) simulations of the CDC composed of gold walls lubricated by hexadecane molecules for a wide range of normal loads and minimum gap heights, which serve as reference data in this work.
The small aspect ratio (i.e. $h_0/L$ in Fig.~\ref{fig:cdc}a) of the CDC motivates modeling the fluid flow as an effective 1D problem, e.g., using the Reynolds equation.
Here, we use the more general height-averaged Navier-Stokes formulation of the fluid problem, which is ideally suited for multiscale simulations of thin film fluid flow~\cite{holey2022_heightaveraged}.

\subsection{Reduced-order continuum model}\label{sec:methods-continuum}
We model the effective fluid flow along the streamwise coordinate ($x$) given a gap topography $h(x)$.
Here, we neglect fluid flow in the spanwise direction ($y$) due to the absence of gap height variations ($\partial h/\partial y=0)$ and the periodicity of the MD system in that direction.
See Ref.~\cite{holey2022_heightaveraged} for a more general formulation of the gap-averaged framework.
The state of the fluid can be described by a gap-averaged (along the third dimension $z$) field
\begin{equation}
\bar{\mathbf{q}}(x, t) = \frac{1}{h(x)} \int_0^{h(x)} \mathbf{q}(x, z, t)\,\dd z,
\end{equation}
where we have assumed that the topography of the lower surface is flat and the upper surface is stationary.
The field $\bar{\mathbf{q}}$ describes an arbitrary collection of conserved variable densities, for which we formulate a continuity equation
\begin{equation}\label{eq:avg_balance}
    \frac{\partial \bar{\mathbf{q}}}{\partial t}  + \frac{\partial \bar{\mathbf{f}}_x}{\partial x} + \mathbf{s} = \mathbf{0},
\end{equation}
where $\bar{\mathbf{f}}_x \equiv \bar{\mathbf{f}}_x(x, t) = h^{-1} \int_{0}^{h} \mathbf{f}_x(x, z, t)\,\dd z $ describes the gap-averaged flux of $\bar{\mathbf{q}}$ in the $x$ direction.
The source term $\mathbf{s}\equiv\mathbf{s}(x, t)$ in Eq.~\eqref{eq:avg_balance} accounts for flux contributions due to the change in topography and fluxes through the bounding walls
\begin{align}\label{eq:src}
\mathbf{s}
&= \frac{1}{h}\left[
 \frac{\dd h}{\dd x}(\bar{\mathbf{f}}_x - \mathbf{f}_x\rvert_{z=h})
+ \mathbf{f}_z\rvert_{z=h} - \mathbf{f}_z\rvert_{z=0}\right].
\end{align}

In this work, we account for mass and momentum conservation under isothermal conditions.
Thus, the relevant state variables are the mass density $\rho$ and the $x$-momentum density (or mass flux) $j_x$, i.e.
\begin{equation}
\mathbf{q}=
 \begin{pmatrix}
 \rho \\
 j_x
\end{pmatrix},
\end{equation}
with the corresponding fluxes
\begin{equation}\label{eq:flux}
 \mathbf{f}_x=
 \begin{pmatrix}
 j_x \\
 j_x^2/\rho - \sigma_{xx}
 \end{pmatrix},
 \qquad
\mathbf{f}_z =
\begin{pmatrix}
 j_z \\
 j_x j_z/\rho - \sigma_{xz}
 \end{pmatrix},
\end{equation}
where $\sigma_{ij}$ denotes a component of the three-dimensional stress tensor $\underline{\sigma}$.
Note that $\mathbf{f}_z$ contains mass flux contributions across the gap $j_z$, but these vanish as we assume impermeable walls ($j_z\rvert_{z\in[0, h]}=0$).
Furthermore, the streamwise flux contains a nonlinear convective term, which is usually small in confined viscous flows.
Given a constitutive model for the relevant stress components $\sigma_{xx}$ and $\sigma_{xz}$ in Eq.~\eqref{eq:flux}, we can solve Eq.~\eqref{eq:avg_balance} numerically.
Here, we used a finite difference implementation on a staggered grid with an explicit time integration algorithm \cite{maccormack2003_effect}.
For more details on the numerical implementation, see Ref.~\cite{holey2022_heightaveraged,huber2026_gapflow}.

\subsection{Data-driven constitutive models using Gaussian process regression}\label{sec:methods-gp}

The solution of Eq.~\eqref{eq:avg_balance} requires the introduction of constitutive models that relate kinematic (e.g. shear rate) or thermodynamic (e.g. density) state variables to the fluid stress,
\begin{equation}
    \sigma_{ij} = f^{(ij)}_\theta(\mathbf{x}),
\end{equation}
where $\mathbf{x}$ denotes a collection of state variables and the subscript $\theta$ denotes a set of parameters (e.g. viscosities, compressibilities) typically obtained from fits to experimental or simulation data.
The superscript $(ij)$ indicates that each stress component is represented by its own function.
While these semi-empirical models are usually grounded in physical principles, they can become difficult to construct when the material behavior is complex, as in boundary lubrication.
Data-driven approaches offer an alternative, in which one seeks a model flexible enough to reproduce a training dataset $\mathcal{D}$ without prior assumptions about its structure.
Here, we focus on probabilistic data-driven models, in which the stress prediction at a test input $\mathbf{x}^\ast$ is expressed as a probability distribution $P$, conditioned on the training data,
\begin{equation}
    \sigma_{ij} \sim P(f^{(ij)}(\mathbf{x}^\ast)\,|\,\mathcal{D}^{(ij)}),
\end{equation}
where the tilde ($\sim$) means ``is distributed according to'', and $\mathcal{D}^{(ij)}$ denotes the training dataset for the stress component $\sigma_{ij}$.

One such model is Gaussian process (GP) regression \citep{rasmussen2006_gaussian}.
In the following, we describe the regression of a single scalar stress component and drop the superscript $(ij)$.
GPs are generalizations of multivariate normal distributions to function spaces and are defined through their mean function $\mu(\mathbf{x})$ and covariance function $k(\mathbf{x}, \mathbf{x}')$, i.e.,
\begin{equation}
    f(\mathbf{x}) \sim \mathcal{GP}\left(\mu(\mathbf{x}), k(\mathbf{x}, \mathbf{x}')\right).
\end{equation}
The one-dimensional flow simulations require predictions of the normal and shear stress components $\sigma_{xx}$ and $\sigma_{xz}$, respectively.
Borrowing ideas from semi-empirical models, we split the normal stress into a viscous part $\tau_{xx}$ and the thermodynamic pressure $p$.
Furthermore, we assume that the viscous part is small, which follows from the thin film assumption in lubricated contacts, i.e. $\sigma_{xx}=\tau_{xx} - p \approx -p$.
Thus, in the following, we use $p$ and $\tau\equiv\sigma_{xz}$ to denote pressure and shear stress in the system, respectively, and reserve the symbol $\sigma$ for GP uncertainties.

We use the Matérn3/2 covariance function \cite{matern1986_spatial}
\begin{equation}
k(\mathbf{x}, \mathbf{x}') = \sigma^2\left(1+\sqrt{3}d(\mathbf{x}, \mathbf{x}')\right)\exp\left(-\sqrt{3}d(\mathbf{x}, \mathbf{x}')\right),
\end{equation}
where $d(\mathbf{x}, \mathbf{x}')=[\sum_{m=1}^{D} (x_m - x'_m)^2/\ell_m^2]^{1/2}$ is the scaled distance between the $D$-dimensional inputs $\mathbf{x}$ and $\mathbf{x}'$ with components $x_m$ and $x'_m$.
The covariance function determines the smoothness of the samples drawn from the GP with its hyperparameters $\sigma$ and $\ell_m$, which describe the amplitude and correlation lengthscale of the covariance.
For convenience, we set the mean function to zero.
To make predictions, we condition the GP on training data.
We denote the training data with $\mathcal{D}=\{(\mathbf{x}_n, y_n)\,|\,n=1,\ldots, N\}$, where $y_n$ denotes the scalar output observation corresponding to input $\mathbf{x}_n$.
The inputs can be further collected into a design matrix $X\in\mathbb{R}^{D\times N}$, and we denote pairwise evaluations of the covariance functions among inputs (i.e. columns in $X$) with $K(X, X)$, leading to an $N\times N$ covariance matrix.
Bayesian inference leads to the predictive distribution evaluated at the test inputs $X^\ast\in\mathbb{R}^{D\times N^\ast}$ given by a multivariate normal distribution
\begin{equation}
    \mathbf{f}^\ast|X,\mathbf{y},X^\ast \sim \mathcal{N}\left(\langle\mathbf{f}^\ast\rangle, \mathrm{cov}\left(\mathbf{f}^\ast\right)\right),
\end{equation}
with the conditioned mean and covariance
\begin{subequations}
\begin{align}
\langle\mathbf{f}^\ast\rangle &= \mu(X^\ast) + K(X^\ast, X) \left[K(X, X) + \sigma_\mathrm{n}^2 I\right]^{-1} [\mathbf{y} - \mu(X)], \\
\mathrm{cov}(\mathbf{f}^\ast) &= K(X^\ast, X^\ast) - K(X^\ast, X) \left[K(X, X) + \sigma_\mathrm{n}^2 I\right]^{-1} K(X, X^\ast),
\end{align}
\end{subequations}
respectively.
We optimize the prediction by maximizing the logarithmic marginal likelihood 
\begin{equation}
\log\, P(\mathbf{y} | X) = -\frac{1}{2}\mathbf{y}^{\top}[K(X,X) +\sigma_\mathrm{n}^2 I]^{-1}\mathbf{y} -\frac{1}{2}\log|K(X,X) +\sigma_\mathrm{n}^2 I| - \frac{N}{2} \log 2\pi
\end{equation}
with respect to the hyperparameters $\sigma$ and $\ell_m$.
The predictive distribution accounts for noisy training data through the noise variance $\sigma_\mathrm{n}^2$, which can be either used as an additional trainable hyperparameter, or is known from the data acquisition process.

We probe the normal and shear stress components as wall tractions in MD simulations (see Sec.~\ref{sec:methods-md}).
Hence, we obtain two stress observations for each of the components: one from the bottom and one from the top wall.
Since pressure is only required in a gap-averaged sense, we take the average of the top and bottom wall component.
The shear stress needs to be evaluated separately for the bottom and top wall, but can be predicted simultaneously using a multi-output GP with the same set of kernel hyperparameters.
Thus, we use two separate GP models for pressure and shear stress.
Both models use inputs composed of the mass density $\rho$, the gap height $h$, and the gradient of the gap topography $\dd h/\dd x$, which influences the structure of the fluid wall interface at the top wall (see Sec.~\ref{sec:app-surface}).
The shear stress model additionally depends on the mass flux $j_x$, accounting for both shear and pressure-driven contributions.

The variance of the predictive distribution can be used for active learning, where the uncertainty guides the acquisition of new MD training data during a multiscale simulation \cite{holey2025_active}.
This strategy can be used to fine-tune an already existing training dataset, or to build one from scratch.
Here, we started with a small dataset of 16 training simulations, and successively augmented the database by running active learning simulations across the training cases (three minimum gap heights and six normal loads).
After two sweeps across the 18 cases, the database contained 135 training points.
For more details on the active learning simulations, we refer to the supplementary material Sec.~\ref{sec:supp_training}.

\subsection{Molecular dynamics training simulations}\label{sec:methods-md}

The training database consists of MD simulations of $n$-hexadecane (\ce{C16H34}) confined between crystalline gold walls.
We closely followed the simulation setup described in Ref.~\citep{codrignani2023_continuum} whose full-scale MD simulations serve as a benchmark for our multiscale simulations.
We used the Transferable Potential for Phase Equilibria in a united atom description (TraPPE-UA) \citep{martin1998_transferable} to model hexadecane, a coarse-grained force field where methyl (\ce{CH3}) and methylene (\ce{CH2}) groups are lumped into representative pseudo atoms.
Gold atoms within the walls interact via the embedded atom method (EAM) potential by \citet{foiles1986_embeddedatommethod}, and the Lennard-Jones parameters for gold by \citet{heinz2008_accurate} were used to model wall fluid interactions via the Lorentz-Berthelot mixing rules.
All MD simulations were performed with LAMMPS \citep{thompson2022_lammps} employing the velocity-Verlet time integration scheme with a time step of $1\,\mathrm{fs}$.

We considered three different MD system setups to account for the characteristic regions in the CDC geometry of Ref.~\citep{codrignani2023_continuum}: one with parallel FCC walls (Fig.~\ref{fig:cdc}c), and two where the crystal orientation of the top wall is rotated around the global $y$ axis (or $[1\bar{1}\bar{2}]$ axis) in either negative (Fig.~\ref{fig:cdc}b) or positive sense (Fig.~\ref{fig:cdc}d).
See Sec.~\ref{sec:app-surface} for details on how the molecular setup with heterogeneous wall structure has been created.
All systems have a lateral box size of $8.65\,\mathrm{nm}$ in $x$-direction and $8.49\,\mathrm{nm}$ in $y$-direction.
Sliding simulations were performed at constant gap height.
Initially, the upper and lower walls were created at a distance larger than the target gap height.
After inserting fluid molecules in the central region between the walls and a short ($25\,\mathrm{ps}$) equilibration period, the walls approached the target gap within $50\,\mathrm{ps}$.
Sliding started after another $25\,\mathrm{ps}$ equilibration period.
During the first $100\,\mathrm{ps}$ without sliding, the temperature of the fluid was controlled with a Berendsen thermostat with a time constant of $0.5\,\mathrm{ps}$.

The outermost layers of the wall slabs are frozen to enforce a constant gap height and to apply the shear deformation, while the four atomic layers closest to the fluid region undergo free dynamics.
We thermalized the remaining four layers in the center of the walls with a Langevin thermostat with a damping time constant $\tau=0.1\,\mathrm{ps}$.
Additionally, we employed a Nosé-Hoover thermostat to the fluid atoms acting only on the velocity degrees of freedom in the spanwise direction ($y$) to ensure isothermal conditions.
In addition to the shear flow introduced by the movement of the lower wall, we constrained the mass flux in the streamwise direction using Gaussian dynamics \citep{strong2017_dynamics}.

The time until a steady state flow has developed in the system depends mainly on the gap height and kinematic viscosity of the fluid.
Due to the wide range of gap heights and viscosities considered in this study, we used a conservative time of $250\,\mathrm{ps}$ to let the systems reach their steady state.
After that, we sampled the pressure and shear stress as wall tractions on the bottom and top walls for another $250\,\mathrm{ps}$ by summing the pairwise forces between fluid and wall atoms in the $z$ (pressure) and $x$ direction (shear stress), and dividing by the surface area $\ell_x\ell_y$.
The standard error of the mean varies slightly between MD systems of different sizes and densities, but for simplicity, we assume homoscedastic noise by setting $\sigma_\mathrm{n}$ to the average value of all training simulations.

\subsection{Reynolds equation}

We compared the results of our multiscale simulations with those based on the Reynolds equation from Ref.~\cite{codrignani2023_continuum}.
The compressible Reynolds equation governing the pressure distribution $p(x)$ in a thin lubricant film of local height $h(x)$ under
steady-state sliding is given by \cite{reynolds1886_iv}
\begin{equation}
    \frac{\dd}{\dd x}\left[\frac{\rho h^{3}}{12\eta}\frac{\dd p}{\dd x}\right]
    =
    \frac{\dd}{\dd x}\left[\rho h\,\frac{v_{x1}+v_{x2}}{2}\right],
    \label{eq:RLE}
\end{equation}
where $\rho$ is the local fluid density, $\eta$ the dynamic viscosity, and $v_{x1}$, $v_{x2}$ are the fluid velocities at the bottom and top wall, respectively. 
When slip is present at the walls, the fluid velocities differ from the wall velocities $v_{x1}^{\mathrm{w}}$ and $v_{x2}^{\mathrm{w}}$ by the local slip velocities $v_{\mathrm{s},1}$ and $v_{\mathrm{s},2}$, i.e.\ $v_{xi} = v_{xi}^{\mathrm{w}} + v_{\mathrm{s},i}$.
Under no-slip conditions one sets $v_{xi}=v_{xi}^{\mathrm{w}}$.
We solve Equation~\eqref{eq:RLE} using a finite difference discretization on a staggered grid.
Because the constitutive models for dynamic viscosity, density (equation of state), and wall slip are nonlinear, we employ a relaxed fixed-point iteration scheme.
The supplementary material (Sec.~\ref{sec:supp_models}) gives an overview of the parametric constitutive models.

\section{Results}

\subsection{Converging-diverging channel simulations}

We created a training dataset by running active learning simulations on the CDC geometry for three minimum gap heights $h_0\in[1.4, 2.4, 4.8]\,\mathrm{nm}$, six normal loads $p_0\approx[0.1, 0.2, 0.4, 0.6, 0.8, 1.0]\,\mathrm{GPa}$, and sliding velocity $v_x^\mathrm{w}=20\,\mathrm{m/s}$.
Note that the target values for normal load are approximate, as we control the average pressure in the continuum simulations by the initial density.
The final training dataset contains 135 MD simulations and has been generated by one forward and one backward sweep across the $6\times 3$ training matrix (see Sec.~\ref{sec:supp_training} for details on the training simulations). 

\begin{figure*}[!ht]
    \centering
    \includegraphics[width=\linewidth]{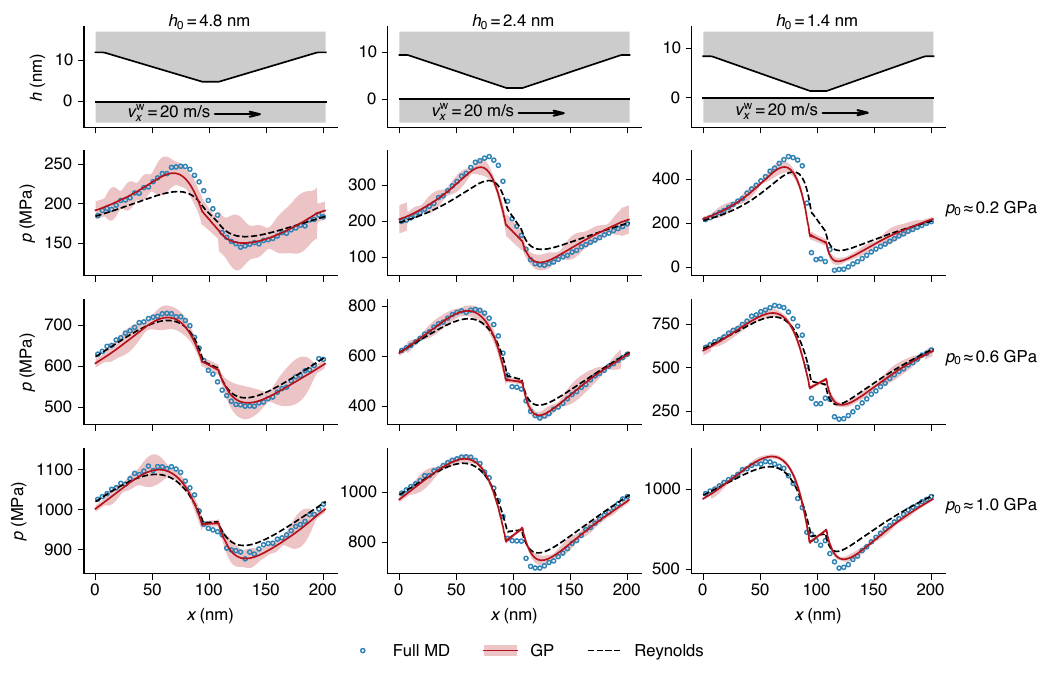}
    \caption{%
        \textbf{Pressure profiles for the converging-diverging channel (CDC) geometry.}
        The first row shows the CDC gap profile for three minimum gap heights $h_0\in[4.8, 2.4, 1.4]\,\mathrm{nm}$ and the rows below show pressure profiles for three normal loads $p_0\in[0.2, 0.6, 1.0]\,\mathrm{GPa}$.
        Molecular dynamics (MD) results from Ref.~\cite{codrignani2023_continuum} are shown as blue circles.
        Pressure profiles obtained by solving the Reynolds equation with the constitutive laws from Ref.~\cite{codrignani2023_continuum} are shown as black dashed lines.
        Solid red lines represent the predictive mean of the Gaussian process (GP) multiscale simulations, while the shaded red areas indicate the 95\% confidence interval given by the GP predictive variance.
        }
    \label{fig:pressure}
\end{figure*}

Figure~\ref{fig:pressure} shows the steady-state pressure profiles for nine selected cases from the full training matrix (see Fig.~\ref{fig:pressure_full}) and compares them to the profiles obtained from large scale MD simulations and solutions to the Reynolds equation using the constitutive models from Ref.~\cite{codrignani2023_continuum}.
The pressure profiles obtained with our multiscale simulations based on GP surrogate models show overall good agreement with the full MD reference.
The GP-based surrogate models show a small pressure increase in the central region for high loads and small gaps (e.g. bottom right corner of Fig.~\ref{fig:pressure}) compared to the lat plateau in the MD and Reynolds simulations.
Note that the predictive uncertainty illustrated by the 95\% confidence interval is similar for all cases but appears larger for the wider gaps due to the different pressure scale.
Compared to the predictions based on the Reynolds equation that work particularly well for large normal loads but fail to reproduce the profiles at low loads and wider gaps, the GP-based simulations work equally well for all normal loads and gap heights.
In nearly all cases, the GP-based simulations predict the pressure minimum right after the transition from the flat to the diverging section of the CDC much better than the Reynolds simulations.

\begin{figure*}[!ht]
    \centering
    \includegraphics[width=\linewidth]{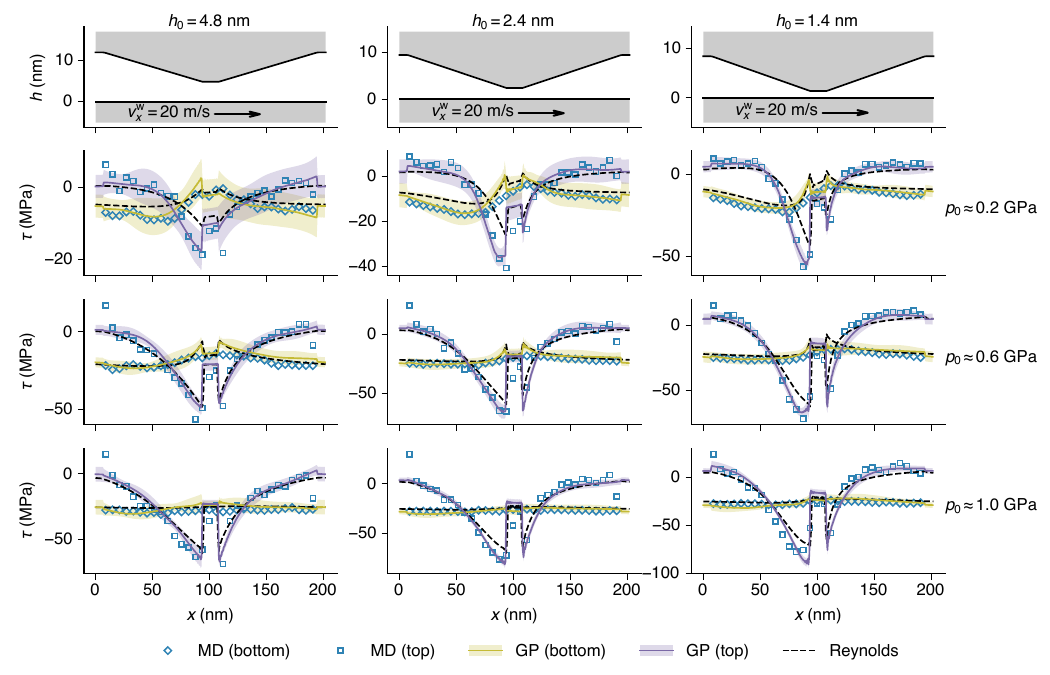}
    \caption{%
        \textbf{Shear stress profiles for the converging-diverging channel (CDC) geometry.}
        The first row shows the CDC gap profile for three minimum gap heights $h_0\in[4.8, 2.4, 1.4]\,\mathrm{nm}$ and the rows below show shear stress profiles at the bottom and top walls for three normal loads $p_0\in[0.2, 0.6, 1.0]\,\mathrm{GPa}$.
        Molecular dynamics (MD) results from Ref.~\cite{codrignani2023_continuum} are shown as blue diamonds and squares for the bottom and top walls, respectively.
        Shear stress profiles obtained by solving the Reynolds equation with the constitutive laws from Ref.~\cite{codrignani2023_continuum} are shown as black dashed lines.
        Solid green and purple lines represent the predictive mean function of the Gaussian process (GP) multiscale simulations for bottom and top walls, respectively.
        The corresponding shaded areas indicate the 95\% confidence interval given by the GP predictive variance.
        }
    \label{fig:shear}
\end{figure*}

Figure~\ref{fig:shear} shows the corresponding steady-state shear stress profiles at the bottom and top wall.
To compare the continuum shear stress based on MD simulations in parallel channels with those measured in the full MD simulations on inclined walls, we need to account for the additional shear stress contribution induced by the fluid pressure.
Thus, for illustration, we corrected the MD shear stress at the top wall
\begin{equation}
    \tau_\mathrm{top, MD} = -\left(\tilde{\tau}_\mathrm{top, MD} + p_\mathrm{MD} \frac{\dd h}{\dd x}\right),
\end{equation}
where $\tilde{\tau}_\mathrm{top, MD}$ denotes the measured shear stress in the full MD simulations. 
The minus sign accounts for the outward pointing surface normal of the top wall used in the measurements of Ref.~\cite{codrignani2023_continuum} which is antiparallel to the global $z$-axis.
The shear stress profiles at both top and bottom surfaces agree closely with those measured in the full scale MD simulations.
The top surface profiles clearly reproduce the jump to a lower stress level when going from an inclined (stepped) to a flat (111) surface in the center of the channels and vice versa.
Bottom stress profiles show similar discontinuities at these transitions, although they are not visible in the full MD results.
As these jumps are less pronounced than those observed on the top profiles, they might have been averaged out during sampling, while the geometric transition is sharply resolved in the continuum simulations.
The improvement compared to the Reynolds calculations observed in Fig.~\ref{fig:pressure} directly corresponds to the shear stress predictions in Fig.~\ref{fig:shear}.
The shear stress profiles for the full training matrix are shown in the supplementary material (Fig.~\ref{fig:shear_full}).

\begin{figure*}[!ht]
    \centering
    \includegraphics[width=\linewidth]{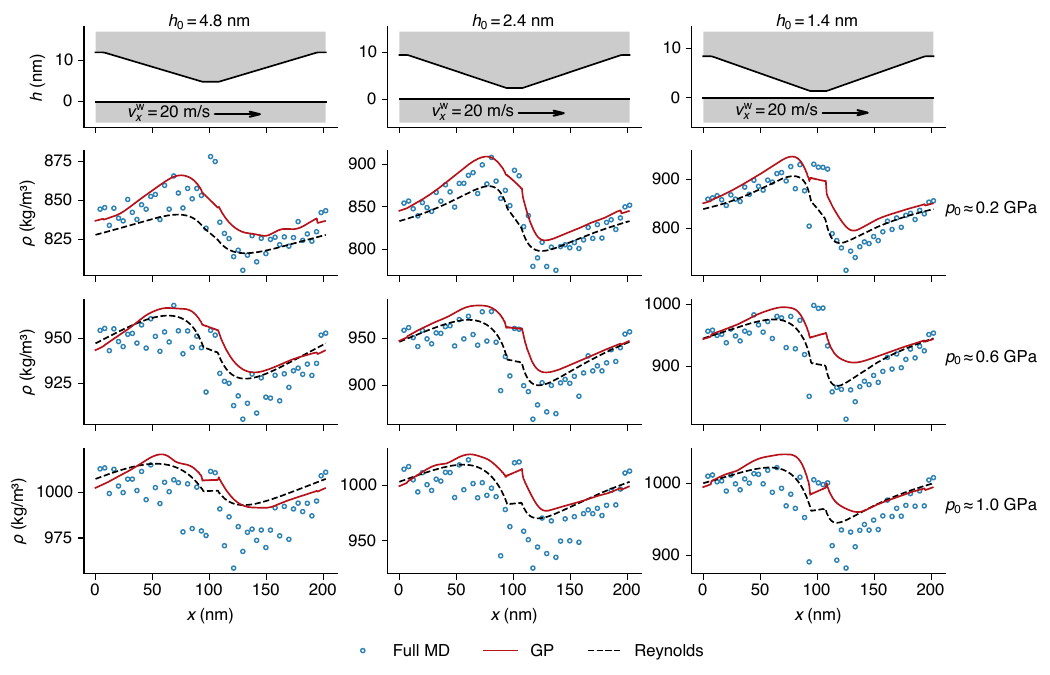}
    \caption{%
        \textbf{Density profiles for the converging-diverging channel (CDC) geometry.}
        The first row shows the CDC gap profile for three minimum gap heights $h_0\in[4.8, 2.4, 1.4]\,\mathrm{nm}$ and the rows below show density profiles for three normal loads $p_0\in[0.2, 0.6, 1.0]\,\mathrm{GPa}$.
        Molecular dynamics (MD) results from Ref.~\cite{codrignani2023_continuum} are shown as blue circles.
        Density profiles obtained by solving the Reynolds equation with the constitutive laws from Ref.~\cite{codrignani2023_continuum} are shown as black dashed lines.
        Solid red lines represent the density profiles obtained with the Gaussian process (GP) multiscale simulations.
        }
    \label{fig:density}
\end{figure*}

Finally, we also compare the steady-state density profiles of our multiscale simulations with those measured in MD, and those obtained from the Reynolds pressure profiles in Fig.~\ref{fig:density} (note that the Reynolds equation is solved for the pressure---a primitive variable---whereas our framework works with conserved variables such as mass density).
The GP-based profiles slightly overestimate the density in all cases compared to the full MD reference and the Reynolds calculations but relative errors remain on the order of a few percent compared to the much larger deviations (up to 100\%) between Reynolds and MD for pressure and shear stress.
Yet, the shape of the profiles matches the MD results well, and some cases even reproduce the visible density jump of the MD profiles.
While we expect some visible density transition when going from regions with rough (stepped) to smooth (111) top wall surfaces due to different layering characteristics  near the surface (compare density profiles in Fig.~\ref{fig:cdc}b and d with those in Fig.~\ref{fig:cdc}c), the extreme jumps observed in some MD simulations are most likely a measuring artifact due to different gap height definitions in the smooth and rough sections.
The density profiles for the full training matrix are shown in the supplementary material (Fig.~\ref{fig:density_full}).

\subsection{Comparison to parametric constitutive laws}

\begin{figure*}[!ht]
    \centering
    \includegraphics[width=\linewidth]{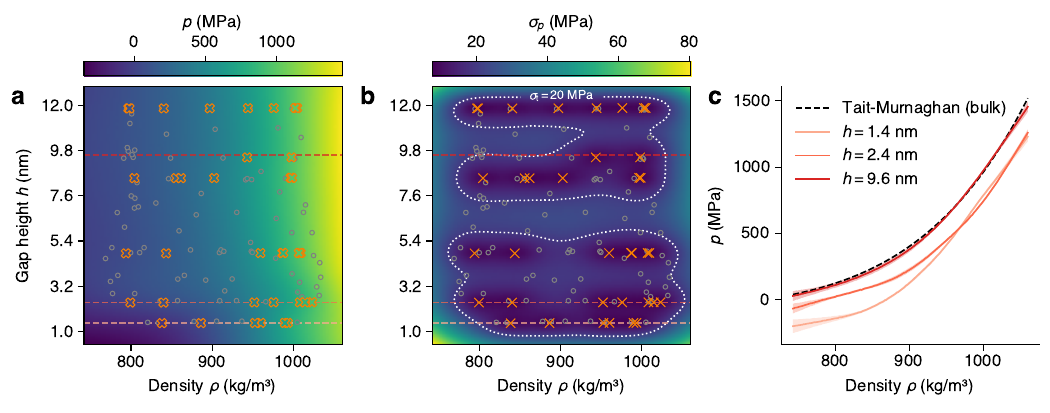}
    \caption{%
        \textbf{Equation of state for flat channels.}
        \textbf{a} Two-dimensional slice through the three-dimensional input space at $\dd h/\dd x=0$, i.e. for flat channels in the widest and narrowest sections of the CDC. Training data at these conditions is highlighted as open orange crosses. The color filling of the symbols corresponds to the measured MD pressure. Training data from non-flat channels are shown as gray circles.
        \textbf{b} Uncertainty of the pressure model illustrated as the standard deviation $\sigma_p$. Orange and gray symbols correspond to the same training data as in \textbf{a}. The white dotted line highlights the boundary between ``accepted'' and ``non-accepted'' regions in an exemplary active learning simulation with uncertainty tolerance $\sigma_\mathrm{t}=20\,\mathrm{MPa}$. 
        \textbf{c} Equation of state at constant gap heights illustrated by the horizontal dashed lines in \textbf{a} and \textbf{b}. The black dashed line highlights the Tait-Murnaghan equation of state as parametrized in Ref.~\cite{codrignani2023_continuum} with bulk MD simulations.
    }
    \label{fig:eos}
\end{figure*}

The GP surrogate model for pressure takes the form $p\equiv p(\rho, h, \dd h/\dd x)$.
Thus, in addition to the density dependence, the equation of state also depends on the gap height and its gradient, which we use as a proxy for surface roughness here.
Figure~\ref{fig:eos}a shows a two-dimensional slice through the three-dimensional input space at constant gradient $\dd h/\dd x=0$.
A similar pressure map is shown in the supplementary material for $\dd h/\dd x<0$ (Fig.~\ref{fig:supp_eos_converging}). 
Thus, it illustrates the equation of state parametrized by MD simulations with ideal (111) bottom and top surfaces as shown in Fig.~\ref{fig:cdc}c.
Open orange cross symbols illustrate the location of the training data in the $\rho$--$h$ plane.
Training data at nonzero gradients, which lie outside this plane, are shown as gray circles.
The pressure model shows the expected monotonic increase with density and only a weak gap height dependence toward smaller heights.

The uncertainty of the pressure prediction is shown in Fig.~\ref{fig:eos}b through the standard deviation $\sigma_p$.
Orange cross and gray circle markers correspond to the training locations inside and outside of the $\dd h/\dd x=0$ slice, respectively.
The overall uncertainty level is low, with a few tens of MPa standard deviation compared to pressure values up to $1\,\mathrm{GPa}$, given the prediction is based on the full training set.
The data within the slice (orange crosses) is clustered at the six available ``flat'' gap heights.
To illustrate the active learning procedure, we chose an arbitrary uncertainty threshold $\sigma_\mathrm{t}=20\,\mathrm{MPa}$ (dotted white line) to distinguish trustworthy from untrustworthy regions of the input space.

Taking 1D slices at constant gap height allows a comparison with the equation of state parametrized in Ref.~\cite{codrignani2023_continuum}.
We compared three different gap heights $h\in[1.4, 2.4, 9.6]\,\mathrm{nm}$ which are shown as dashed horizontal lines in Fig.~\ref{fig:eos}a and b.
The pressure-density relation at the largest gap height is almost identical with the Tait-Murnaghan \cite{murnaghan1944_compressibility} fit to bulk MD data from Ref.~\cite{codrignani2023_continuum}.
The pressure is lower for the two smaller gap heights than in the bulk-like case, being nearly identical for densities above $950\,\mathrm{kg/m^3}$.
At lower densities, the pressure at $1.4\,\mathrm{nm}$ is much lower than at $2.4\,\mathrm{nm}$, falling even below zero at approximately $880\,\mathrm{kg/m^3}$.
Comparison with Figs.~\ref{fig:density} and \ref{fig:density_full}, however, shows that the steady-state densities in the narrowest region of the CDC barely fall below $900\,\mathrm{kg/m^3}$.
Nevertheless, the CDC with a minimum gap height of $1.4\,\mathrm{nm}$ at the lowest normal load of $0.1\,\mathrm{GPa}$ seems to be right at the edge of cavitation (see Fig.~\ref{fig:pressure_full}).

In addition to density, gap height, and the height gradient, the shear stress model also depends on the mass flux $j_x$.
For a direct comparison with the parametric models from \citet{codrignani2023_continuum}, we take a slice at constant density $\rho=941\,\mathrm{kg/m^3}$ and constant mass flux $j_x=\rho v_x^\mathrm{w} / 2$, which corresponds to ideal Couette flow under no-slip conditions.
The wall velocity $v_x^\mathrm{w}$ is not part of the input space dimensions, since all training simulations were performed at constant wall velocity $v_x^\mathrm{w}=20\,\mathrm{m/s}$.
Yet, when plotted over the applied shear rate $\dot{\gamma} = v_x^\mathrm{w}/h$, we can compare the nonparametric shear stress model directly with the parametric model from Ref.~\cite{codrignani2023_continuum} under the same conditions.
Note that the parametric shear stress prediction shows the combined effect of pressure-dependent viscosity, shear-thinning, and wall slip models as presented in Sec.~\ref{sec:supp_models}.

\begin{figure*}[!ht]
    \centering
    \includegraphics[width=\linewidth]{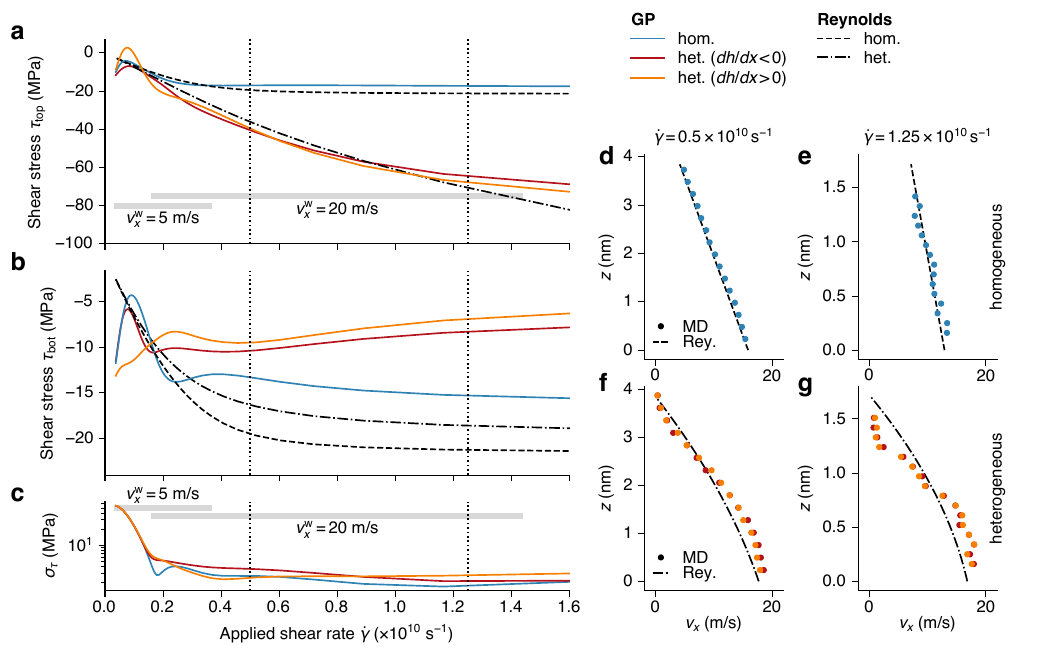}
    \caption{%
        \textbf{Comparison of nonparametric and parametric shear stress models.}
        \textbf{a} Shear stress at the top wall as a function of the applied shear rate. GP-based models are generated as slices at $\rho=941\,\mathrm{kg/m^3}$ and $j_x=v_x^\mathrm{w}\rho/2$ which corresponds to perfect Couette-type flow under no-slip conditions. The results for the three different MD systems are shown as blue (homogeneous walls $\dd h/\dd x=0$), red (heterogeneous, converging walls $\dd h/\dd x<0$), and orange (heterogeneous, diverging walls $\dd h/\dd x>0$) lines. The wall slip constitutive law from Ref.~\cite{codrignani2023_continuum} does not distinguish between converging and diverging sections in the heterogeneous systems. Thus, the dashed and dash-dotted black lines correspond to homogeneous and heterogeneous systems, respectively.
        \textbf{b} Same as \textbf{a} but for the bottom wall.
        \textbf{c} Predictive uncertainty $\sigma_\tau$ of the GP-based models. The two vertical dotted lines in panels \textbf{a}--\textbf{c} indicate shear rates at which we directly compare MD velocity profiles with those predicted by the parametric models and the Reynolds equation.
        \textbf{d} MD velocity profiles at applied shear rate $\dot{\gamma}=5\times10^9\,\mathrm{s}^{-1}$ in the homogeneous system (blue disks) and comparison to the Reynolds prediction (black dashed line).
        \textbf{e} Same as \textbf{d} but for $\dot{\gamma}=1.25\times10^{10}\,\mathrm{s}^{-1}$.
        \textbf{f} MD velocity profiles at applied shear rate $\dot{\gamma}=5\times10^9\,\mathrm{s}^{-1}$ in the heterogeneous system (red and orange disks for converging and diverging sections, respectively) and comparison to the Reynolds prediction (black dash-dotted line).
        \textbf{g} Same as \textbf{f} but for $\dot{\gamma}=1.25\times10^{10}\,\mathrm{s}^{-1}$.
    }
    \label{fig:shear_model}
\end{figure*}

Figure~\ref{fig:shear_model}a shows the shear stress at the top wall for all three MD channel types (homogeneous, heterogeneous converging, heterogeneous diverging) and the prediction from the parametric models, which do not distinguish between converging and diverging sections in the heterogeneous channels.
In the range of training simulations at $v_x^\mathrm{w}=20\,\mathrm{m/s}$, as indicated by the horizontal gray bar, the nonparametric and parametric models agree well with each other with some small deviations for heterogeneous channels at high shear rates. 
In the homogeneous channels, the magnitude of the top wall shear stress quickly saturates at around $20\,\mathrm{MPa}$, while in the heterogeneous channels with a rough top wall, the top wall shear stress continues to grow with shear rate.
At high shear rates ($>10^{10}\,\mathrm{s}^{-1}$) the GP model's prediction levels off as well, but at higher stress magnitudes than in the homogeneous channels.
Note that \citet{codrignani2023_continuum} assume no-slip conditions at the top wall of heterogeneous channels such that the only possible mechanism of shear stress reduction is due to shear thinning.

On the smooth bottom surfaces, the overall shear stress level is low for all channel types and models (see Fig.~\ref{fig:shear_model}b) with a maximum shear stress magnitude around $20\,\mathrm{MPa}$.
All models predict nearly constant shear stress levels in the range of the applied shear rates of the training simulations.
Most notable is the slight decrease of the shear stress level with increasing shear rate for the heterogeneous channels in the nonparametric models.
The shear stress prediction of the parametric model for the homogeneous channel is equivalent to the top surface for Couette-type flows with equal slip on both walls, while heterogeneous channels have slightly lower stress levels.
The nonparametric GP models predict lower shear stress magnitudes for both channel types.
The difference between parametric and nonparametric models is more pronounced for heterogeneous channels and high shear rates, reaching approximately $10\,\mathrm{MPa}$.

Figure~\ref{fig:shear_model}c shows the predictive uncertainty of the GP models.
The uncertainty given by the standard deviation $\sigma_\tau$ is less than $5\,\mathrm{MPa}$ in the range of training simulations at $v_x^\mathrm{w}=20\,\mathrm{m/s}$.
The applied shear rates at $v_x^\mathrm{w}=5\,\mathrm{m/s}$ reach into a regime where the standard deviation is larger than $10\,\mathrm{MPa}$, and thus, larger than the predictive mean.
At applied shear rates below $10^9\,\mathrm{s}^{-1}$, the GP model for these particular slices through the four-dimensional input space is therefore unreliable.

Velocity profiles reveal the actual modes of shear accommodation and may explain some of the differences between the parametric and nonparametric shear stress models.
We highlight two selected shear rates with vertical dotted lines in Fig.~\ref{fig:shear_model}a-c, for which we conducted MD simulations for all three channel types and compared the velocity profiles with those predicted by the Reynolds solver.
In homogeneous channels under Couette-type flow, we expect wall slip to be equal on the opposing surfaces which would lead to linear velocity profiles.
This behavior is confirmed for MD and Reynolds velocity profiles for both low shear rate $\dot{\gamma}=5\times10^9\,\mathrm{s}^{-1}$ (with gap height $h=4\,\mathrm{nm}$) in Fig.~\ref{fig:shear_model}d and high shear rate $\dot{\gamma}=1.25\times10^{10}\,\mathrm{s}^{-1}$ (with gap height $h=1.6\,\mathrm{nm}$) in Fig.~\ref{fig:shear_model}e.
Since the magnitude of the slip velocity predicted by the wall slip model of Ref.~\cite{codrignani2023_continuum} agrees well with the MD profiles, the difference in shear stress is likely due to their viscosity model.
Here, it is important to keep in mind that both shear thinning and pressure-dependent viscosity models have been parametrized with bulk MD simulations in Ref.~\cite{codrignani2023_continuum}, while the GP models in this work take wall-induced effects, such as viscosity changes due to layering, into account.

For heterogeneous channels, wall slip at top and bottom surfaces differs, which leads to a departure from the linear velocity profiles in order to maintain the fixed mass flux as shown in Fig.~\ref{fig:shear_model}f and g for the low and high shear rate, respectively.
Although the Reynolds-predicted wall slip at the bottom wall is similar to that measured in MD, the shape of the velocity profiles differs.
Both MD velocity profiles are S-shaped, concentrating some of the shear deformation in regions inside the channel.
The Reynolds calculations from Ref.~\cite{codrignani2023_continuum} cannot reproduce this behavior, which is expected since they do not account for viscosity variations across the gap as in generalized Reynolds solvers (but could easily be extended to do so).
This behavior is more pronounced in the $1.6\,\mathrm{nm}$ thin channel (Fig.~\ref{fig:shear_model}g), where strong layering occurs and shear predominantly happens between fluid layers.
Such behavior is difficult to include in conventional Reynolds solvers, but is automatically captured by our GP models.
Furthermore, there is no substantial difference between the MD velocity profiles in converging and diverging heterogeneous channels (compare red and orange disks in Fig.\ref{fig:shear_model}f and g).

\subsection{Choice of features and transferability}

All MD training simulations and test cases have been performed at wall velocity $v_x^\mathrm{w}=20\,\mathrm{m/s}$.
Thus, although not explicitly used as an input, the GP models are not directly transferable to cases with different wall velocities.
As already pointed out above, the applied shear rate $\dot{\gamma}=v_x^\mathrm{w}/h$ would be a more common independent variable for the shear stress model.
Due to the flexibility of the GP models, an extension to shear rate dependent models is straightforward, which leads to a five-dimensional input space, since we keep the explicit gap height dependence.
This new model is now transferable to cases with different wall velocities.

\begin{figure*}[!ht] 
    \centering
    \includegraphics[width=\linewidth]{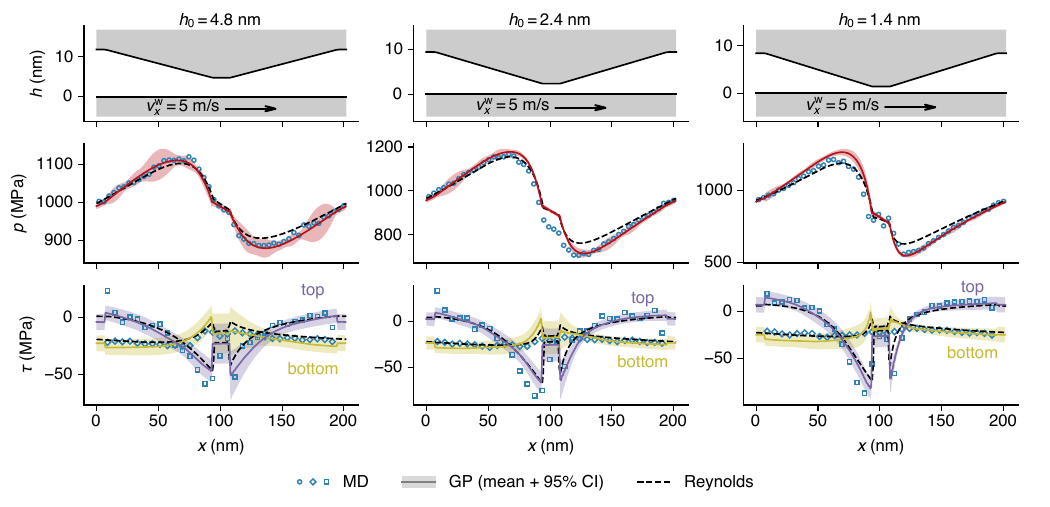}
    \caption{%
        \textbf{Pressure and shear stress prediction for CDC geometries at $v_x^\mathrm{w}=5\,\mathrm{m/s}$ and $p_0\approx1\,\mathrm{GPa}$.}
        The first row shows the CDC gap profile for three minimum gap heights $h_0\in[4.8, 2.4, 1.4]\,\mathrm{nm}$.
        The second row shows the pressure profiles, where solid red lines represent the GP predictive mean, and the red shaded areas indicate the 95\% confidence interval (CI) given by the GP predictive variance.
        The third row shows the GP shear stress prediction and uncertainty at the bottom and top wall in light green and purple, respectively.
        Molecular dynamics (MD) results from Ref.~\cite{codrignani2023_continuum} for pressure, bottom wall shear stress and top wall shear stress are shown as blue circles, diamonds, and squares, respectively.
        Pressure and shear stress profiles obtained by solving the Reynolds equation with the constitutive laws from Ref.~\cite{codrignani2023_continuum} are shown as black dashed lines.
   }
    \label{fig:extrapolate}
\end{figure*}

\citet{codrignani2023_continuum} also conducted atomistic CDC simulations at wall velocity $v_x^\mathrm{w}=5\,\mathrm{m/s}$.
The range of the applied shear rates overlaps with those at wall velocity $v_x^\mathrm{w}=20\,\mathrm{m/s}$ as shown by the gray horizontal bars in Fig.~\ref{fig:shear_model}a and c.
Without adding new training data from MD simulations at $5\,\mathrm{m/s}$, we conducted test simulations on the CDC geometry at the lower wall velocity and compared them to the full atomistic reference in Fig.~\ref{fig:extrapolate}.
The second and third row of Fig.~\ref{fig:extrapolate} show the pressure and shear stress profiles obtained for the largest normal load $p_0\approx 1\,\mathrm{GPa}$, respectively.
The agreement with the MD profile is remarkable for both pressure and shear stress, although some parts of the solution clearly extrapolate into unknown domains of the input space.
The GP uncertainty provides a measure that indicates when this extrapolation becomes unreliable as we have seen in Fig.~\ref{fig:shear_model}c.
This is not the case for the examples shown here, which do not show extremely wide confidence intervals, and where the overall prediction is slightly better than that of the Reynolds equation with parametric constitutive laws.

\section{Discussion}

The GP surrogate models achieve accuracy equal to or better than the Reynolds equation when compared against full-scale MD reference simulations of the CDC.
This result is surprising at first glance, given that the Reynolds equation is supplied with carefully calibrated constitutive laws for compressibility, piezoviscosity, and shear thinning.
However, the parametrization of viscosity laws on bulk representative volume elements and the assumption of constant viscosity across the gap might explain the fixed-form model's difficulties in describing the transition between slip-dominated and viscosity-dominated flow regime across the CDC simulations.
The GP surrogates, by contrast, are trained directly on confined MD systems, and are flexible enough to capture both regimes without distinguishing between them explicitly.
The model adapts continuously to whatever stress response the fluid exhibits at a given set of input conditions.

Fixed-form constitutive models are useful, particularly when they are developed on a non-empirical basis, where much can be learned about the underlying physical mechanisms \cite{falk2020_nonempirical}.
They can achieve similar flexibility, for instance by allowing viscosity or compressibility to depend on gap height \cite{martini2006_molecular,vadakkepatt2011_confined}, or by explicitly including the effect of adsorbed boundary layers \cite{zhang2020_modeling}.
However, these approaches usually come at the cost of additional assumptions and empirical parameters.
Increasing model complexity not only demands more data, but also erodes the models' original advantage of mechanistic interpretability.
Furthermore, the selection of parametric models, and thus mechanistic understanding, is sometimes ambiguous.
For instance, \citet{codrignani2023_continuum} also tested an Eyring-type slip law \cite{eyring1936_viscosity}, which has---in contrast to the Thompson-Troian model \cite{thompson1997_general}---a clear underlying physical picture of stress-assisted and thermally activated jumps of lubricant segments on solid surfaces \cite{lichter2007_liquid,wang2011_slipa, wang2019_universal}.
While both models worked equally well for the gold hexadecane system, the Eyring model described MD data generated for a polyalphaolefin lubricant on diamond-like carbon surfaces better.
\citet{jadhao2019_rheological} found that shear thinning of squalane is best described by the Carreau model at low pressure and by the Eyring model at high pressure, while \citet{codrignani2023_continuum} use the Carreau model for pressures up to $2\,\mathrm{GPa}$ for hexadecane.
The GP models avoid these issues altogether by not committing to any particular mechanism.
The number of MD simulations required to train the GP surrogates is comparable to or smaller than what would be needed to calibrate a sufficiently flexible fixed-form model.
This efficiency is in part due to the active learning algorithm, which selects new training points exclusively in the physically relevant regions of the high-dimensional input space, avoiding the redundant coverage of unvisited regions that characterizes grid-based or design-of-experiment sampling strategies.

We hope that our data-driven multiscale simulations will stimulate a paradigm shift in constitutive modeling for boundary lubrication.
Rather than seeking simple mechanism-based models that isolate individual physical contributions, purely data-driven surrogates account for the combined effect of all active mechanisms simultaneously.
The enhanced accuracy of such approaches comes, however, at the cost of reduced mechanistic interpretability and transferability.
A GP model trained for a specific molecular system and interatomic potential encodes both functional form and material-specific information jointly into its covariance structure.
Unlike fixed-form models, GP surrogates offer no natural route for recalibration against experimental rheological or friction data for real, chemically complex lubricants.
Yet, physical insight is not completely abandoned, since the choice of input features encodes physical assumptions about which variables govern the stress response, and sensitivity analysis with respect to individual inputs can be used to quantify the relative importance of different contributions.

A route toward predictive modeling of arbitrary real world lubricants could be achieved by combining the best of both worlds.
Fixed-form constitutive laws, calibrated with bulk MD and experimental data, could serve as a GP prior mean function, governing the stress response in regions of relatively thick lubricant films and mild conditions.
In regions where fundamental assumptions of the underlying constitutive models are broken and non-continuum effects dominate, the flexible GP model corrects the fixed-form laws using MD data of the confined system.
Realizing this vision at scale would profit from a community-driven approach to constitutive modeling, in which shared databases of MD simulations, organized according to FAIR data principles \cite{wilkinson2016_fair}, are used to train surrogate models of increasing accuracy and transferability.
Such models could close the knowledge gap in the mixed and boundary lubrication regime and would enable physics-guided design of frictional interfaces in modern engineering systems.

\section{Conclusion}

We have presented a nonparametric multiscale framework for boundary lubrication that couples Gaussian process surrogate models to a continuum thin-film solver, bypassing the need for fixed-form constitutive laws. 
Applied to hexadecane confined between gold surfaces under pressures up to $1\,\mathrm{GPa}$ and gap heights down to $1.4\,\mathrm{nm}$, the framework achieves accuracy comparable to or better than a state-of-the-art Reynolds description with carefully calibrated constitutive laws, while requiring a smaller number of molecular dynamics simulations.
The surrogate models naturally capture the transition between slip-dominated and viscosity-dominated flow regimes without explicit mechanistic assumptions, and an active learning algorithm ensures that the training database is populated efficiently and exclusively in physically relevant regions of the input space.
These results demonstrate that nonparametric surrogate models offer a flexible and robust alternative for predictive continuum modeling of boundary lubrication, and suggest a broader paradigm shift from mechanism-based constitutive laws toward data-driven approaches that account for the combined effect of multiple molecular mechanisms in extreme environments.
Future work will focus on extending the framework to more complex lubricant-surface combinations and on driving the systems to even more extreme cases, approaching the limit of a single lubricant layer.

\begin{acknowledgments}
We thank James Kermode for many useful discussions on data-driven modeling. H.H. thanks the Alexander von Humboldt Foundation for support through the Feodor Lynen fellowship.
M.M., acknowledges funding from the European Research Council (ERC) under the European Union’s Horizon 2020 research and innovation programme (ERC Advanced Grant LubeTwin, agreement No. 101201061).
\end{acknowledgments}

\clearpage


\renewcommand{\thefigure}{S\arabic{figure}}    
\setcounter{figure}{0}    
\renewcommand{\thesection}{S\arabic{section}}
\setcounter{section}{0}
\renewcommand{\thetable}{S\arabic{table}}
\setcounter{table}{0}
\renewcommand{\theequation}{S\arabic{equation}}
\setcounter{equation}{0}
\renewcommand{\thepage}{S\arabic{page}}
\setcounter{page}{1}

\begin{center}
\textbf{\large Supplementary Material for:\\ Nonparametric multiscale modeling of boundary lubrication: hexadecane in highly pressurized gold asperity contacts}
\end{center}

\section{Preparation of MD boxes with stepped walls}\label{sec:app-surface}
\begin{figure*}[!ht]
    \centering
    \includegraphics[]{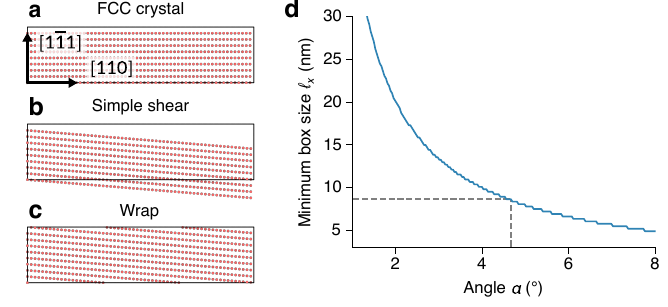}
    \caption{%
    \textbf{Generation of stepped periodic walls.}
    \textbf{a} The perfect FCC crystal with its $[110]$ direction aligned with the streamwise coordinate $x$ and the $[1\bar{1}1]$ direction aligned with the $z$ coordinate.
    \textbf{b} The top walls in regions with $|\alpha|>0$ are created by applying a simple shear deformation corresponding to $\alpha$ (here $\alpha>0$).
    \textbf{c} Atoms that are sheared outside of the initial box are re-mapped into it to create the stepped surface pattern.
    \textbf{d} Minimum box size for a given angle that can be realized without breaking the periodicity. Larger boxes can be created using integer multiples of the minimum size.
    }
    \label{fig:supp_rough_wall}
\end{figure*}

The converging--diverging channel (CDC) geometry of the full scale MD simulations from Ref.~\citep{codrignani2023_continuum} is carved out of an FCC single crystal, where the $(111)$ plane lies within the global $x$--$y$ plane.
This leads to steps on the upper wall surfaces of the converging and diverging sections (see Fig.~\ref{fig:cdc}a of the main text).
Although topography gradients are already considered on the continuum scale, these step-patterned surfaces lead to distinct fluid-wall interactions compared to the flat sections (see Fig.~\ref{fig:cdc}b-d of the main text).
Thus, for an accurate representation of the local flow state in the converging and diverging sections, we created heterogeneous MD systems with an ideal flat bottom surface and a stepped top surface.
We created the top surface by applying a simple shear deformation to the perfect crystal slab (Fig.~\ref{fig:supp_rough_wall}a)  corresponding to $\alpha=\tan^{-1}{(\dd h/\dd x)}$ given by the deformation gradient $\underline{F}=\underline{1}-\tan{\alpha}\ \vec{e}_z \otimes \vec{e}_x$ (Fig.~\ref{fig:supp_rough_wall}b), and mapping the transformed atoms back into the initial rectangular box (Fig.~\ref{fig:supp_rough_wall}c).
Note that the slope of a $(111)$ plane after the transformation has the opposite sign of the topography gradient to create the correct step pattern.
To match the periodicity of the FCC lattice of the lower wall, only certain angles are allowed for a given lateral size ($\ell_x$) of the simulation box (see Fig.~\ref{fig:supp_rough_wall}d) and vice versa.
For the CDC geometry with inclination angles $\pm 4.7$\textdegree, the lateral size should be an integer multiple of $\ell_x=8.65\,\mathrm{nm}$.

\section{Active learning simulations}\label{sec:supp_training}

\begin{figure*}[!ht]
    \centering
    \includegraphics[width=\linewidth]{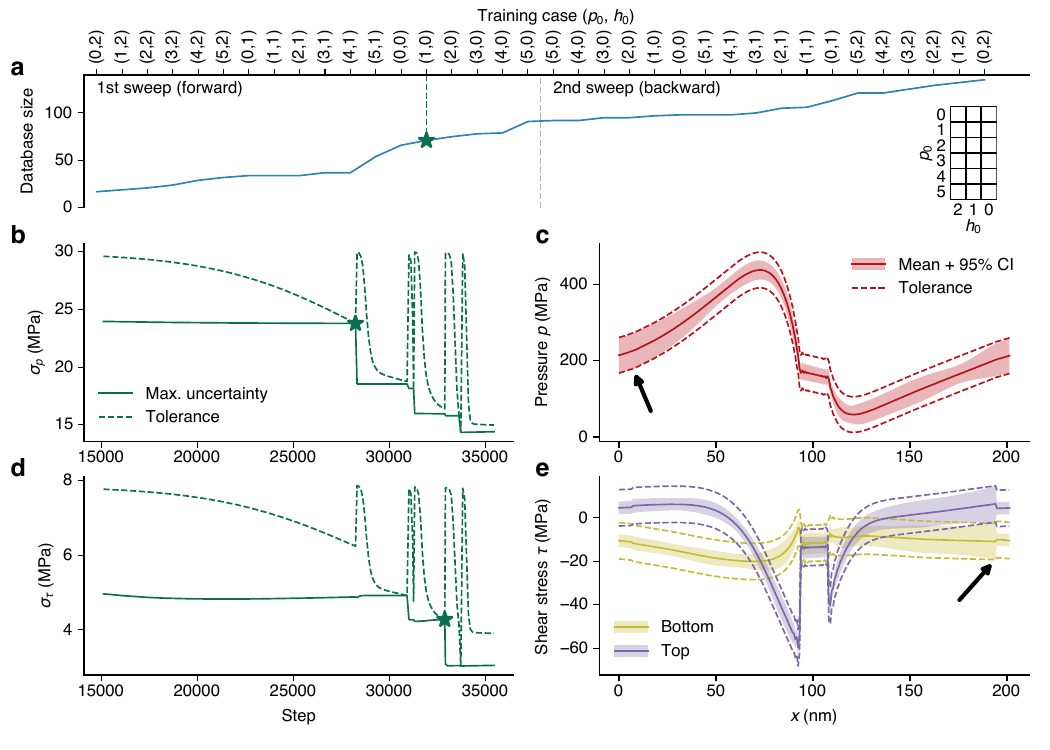}
    \caption{%
    \textbf{Active learning simulations.}
    \textbf{a} Growth of the training database during two sweeps across the training case matrix at $v_x^\mathrm{w}=20\,\mathrm{m/s}$. The first sweep starts at case $(0,\,2)$, i.e., at the smallest normal load $p_0\approx0.1\,\mathrm{GPa}$ and the largest minimum gap height $h_0=4.8\,\mathrm{nm}$. The second sweep reverses the first one by starting at the most extreme case $(5,\,0)$, i.e., at the largest normal load $p_0\approx1.0\,\mathrm{GPa}$ and the smallest minimum gap height $h_0=1.4\,\mathrm{nm}$. 
    \textbf{b} Evolution of the maximum uncertainty of the pressure model given by the standard deviation $\sigma_p$ (solid line) and the adaptive tolerance (dashed line) for case $(1,\,0)$. The star highlights the data acquisition triggered by the pressure model.
    \textbf{c} GP pressure prediction just before the data acquisition. The arrow points to the location where the uncertainty exceeds the adaptive tolerance.
    \textbf{d} Same as \textbf{b} but for the shear stress model.
    \textbf{e} Same as \textbf{c} but for the shear stress model.
    }
    \label{fig:supp_learning}
\end{figure*}

The training database for the GP-based multiscale simulations was generated using active learning \cite{holey2025_active}.
To this end, we arranged 18 training cases at six different normal loads $p_0\approx[0.1, 0.2, 0.4, 0.6, 0.8, 1.0]\,\mathrm{GPa}$ and three different minimum gap heights $h_0\in[1.4, 2.4, 4.8]\,\mathrm{nm}$ into a $6\times3$ matrix.
All training simulations were conducted at wall velocity $v_x^\mathrm{w}=20\,\mathrm{m/s}$ and the evolution of the database size is shown in Fig.~\ref{fig:supp_learning}a.
We started with a small initial database of 16 MD simulations and a case which corresponds to the mildest loading conditions with the largest minimum gap height $h_0=4.8\,\mathrm{nm}$ and the smallest normal load $p_0\approx0.1\,\mathrm{GPa}$, i.e. at case $(0,\,2)$.
During a first sweep, we ran simulations for all loading conditions in the matrix in a row-major fashion, i.e. increasing the normal load before increasing the minimum gap height.
After the first sweep, we reversed the direction starting from the most severe loading conditions with the largest normal load $p_0\approx1\,\mathrm{GPa}$ and the smallest minimum gap $h_0=1.4\,\mathrm{nm}$ (i.e. at case $(5,\,0)$).

During an active learning simulation of a single case, we augment the training database whenever the GP predictive uncertainty exceeds a predefined tolerance.
As an example, Fig.~\ref{fig:supp_learning}b shows the evolution of the maximum uncertainty of the pressure model for case $(1,\,0)$ of the first sweep as highlighted by a star in Fig.~\ref{fig:supp_learning}a.
We used an adaptive uncertainty tolerance, which reduces the tolerance level as the solution approaches a steady state.
The adaptive tolerance is defined as a sigmoid function
\begin{equation}
    \sigma_\mathrm{t}(s) = \sigma_\mathrm{t}^\infty + 
    \frac{\sigma_\mathrm{t}^0 - \sigma_\mathrm{t}^\infty}{1 + \exp(-s)},
\end{equation}
where $\sigma_\mathrm{t}^0$ and $\sigma_\mathrm{t}^\infty$ denote the initial and final uncertainty tolerance, respectively, and we parametrize $s$ with the residual $r$ of the transient numerical solution, $s(r)=\beta\ln(r/r_0)$.
Thus, the parameters $r_0$ and $\beta$ control when and how fast the tolerance reduction takes place, respectively.
For all simulations, we used $r_0=10^{-5}$ and $\beta=2$.
For both models, we used an initial tolerance level $\sigma_\mathrm{t}^0=5\sigma_\mathrm{n}$, where $\sigma_\mathrm{n}$ is the measurement error from MD simulations, and a final tolerance $\sigma_\mathrm{t}^\infty=\sigma_\mathrm{t}^0/2$.
For the pressure model, this leads to an initial tolerance of approximately $30\,\mathrm{MPa}$.
The parametrization with the residual has the advantage that the tolerance falls back to a high level when the addition of new data leads to sudden changes in the numerical solution due to changes of the constitutive model.
This can be seen from the tolerance jumps at data acquisition steps, where the maximum uncertainty and the tolerance coincide, such as the one highlighted with a star in Fig.~\ref{fig:supp_learning}b.

Figure~\ref{fig:supp_learning}c shows the pressure prediction at the time step just before new data is added.
The uncertainty and tolerance are shown as a 95\% confidence interval around the predictive mean.
The arrow indicates the location along the streamwise coordinate $x$ of the CDC, that triggered the addition of new data.

The addition of new data triggered by the pressure model may also affect the shear stress model, depending on the location of the new data point.
Figure~\ref{fig:supp_learning}d shows the time evolution of maximum uncertainty and tolerance for the shear stress model.
Note that the multi-output GP model for the shear stress at the bottom and top wall provides only a single uncertainty.
The initial tolerance is approximately $8\,\mathrm{MPa}$ and the adaptive tolerance jumps at the same time frames as in the pressure model.
Similar to the pressure model, we highlight a case where the shear stress model triggered data acquisition with a star.
The shear stress prediction at the time step just before the active learning step is shown in Fig.~\ref{fig:supp_learning}e and the arrow indicates the location where the uncertainty reaches the maximum allowable value.

\section{Parametric constitutive models}\label{sec:supp_models}

Here, we provide a short overview of the constitutive models and the parameters used in Ref.~\cite{codrignani2023_continuum}.
%
The pressure dependence of the fluid density is described by the Tait--Murnaghan~\cite{murnaghan1944_compressibility} equation of state
\begin{equation}
    \rho(p) = \rho_{0}
    \left[1 + \frac{n_\mathrm{TM}}{K_\mathrm{TM}}
    \left(p - p_\mathrm{TM}\right)\right]^{1/n_\mathrm{TM}},
    \label{eq:TaitMurnaghan}
\end{equation}
where $\rho_{0}$ is the reference density, $K_\mathrm{TM}$ the bulk modulus, $p_\mathrm{TM}$ the reference pressure, and $n_\mathrm{TM}$ a dimensionless exponent.

The viscosity depends on both pressure $p$ and local shear rate $\dot{\gamma}$.  
The Newtonian (zero-shear) viscosity at elevated pressure is captured by the Roelands piezoviscosity formula~\cite{roelands1966_correlational}
\begin{equation}
    \eta_{\mathrm{N}}(p) = \eta_{0}
    \exp\!\left\{
        \ln\!\left(\frac{\eta_{0}}{\eta_{\infty}}\right)
        \!\left[-1 + \left(1 + \frac{p}{p_{\mathrm{R}}}\right)^{z_{\mathrm{R}}}\right]
    \right\},
    \label{eq:Roelands}
\end{equation}
with ambient viscosity $\eta_{0}$, high-pressure limiting viscosity $\eta_{\infty}$, pressure scale $p_{\mathrm{R}}$, and Roelands exponent $z_{\mathrm{R}}$.
Shear thinning is incorporated via the Carreau model~\cite{carreau1972_rheological}
\begin{equation}
    \eta(\dot{\gamma}, p) = \eta_{\mathrm{N}}(p)
    \left[1 + \left(\frac{\dot{\gamma}}{\dot{\gamma}_{0}(p)}\right)^{2}
    \right]^{\!\frac{n_\mathrm{Car}(p)-1}{2}},
    \label{eq:Carreau}
\end{equation}
where $\dot{\gamma}_{0}(p)$ is the pressure-dependent characteristic shear rate and $n_\mathrm{Car}(p)$ the Carreau power-law exponent. Both parameters depend exponentially on pressure:

\begin{equation}
    \dot{\gamma}_{0}(p) = \dot{\gamma}_{00} \exp(-p/p_{\dot{\gamma}_0}), \qquad
    n_\mathrm{Car}(p) = n_0 \exp(-p/p_{n_\mathrm{Car}}).
    \label{eq:Carreau_params}
\end{equation}

When the no-slip boundary condition is violated, the slip velocity $v_{\mathrm{s}}$ at a wall is related to the local shear stress $\tau$ and pressure $p$ through a
constitutive slip law. \citet{codrignani2023_continuum} employed two formulations: the Thompson--Troian slip law~\cite{thompson1997_general}, and the Eyring slip law based on molecular kinetic theory.
Here, we consider only the Thompson--Troian relation, as the Eyring slip law led to indistinguishable results for the gold-hexadecane system in Ref.~\cite{codrignani2023_continuum}.
Thus, the slip velocity reads
\begin{equation}
    v_{\mathrm{s}}(\tau, p) = \frac{v_{\mathrm{c}}(p)}{2}
    \sqrt{\frac{\tau^{2}}{\tau_{\mathrm{c}}(p)\left[\tau_{\mathrm{c}}(p)-\tau\right]}},
    \label{eq:TT}
\end{equation}
where $\tau_{\mathrm{c}}(p)$ is the limiting (maximum) shear stress the fluid--wall interface can sustain, and $v_{\mathrm{c}}(p)$ is a characteristic slip velocity.
Both parameters depend linearly and exponentially on pressure, respectively:
\begin{equation}
    \tau_{\mathrm{c}}(p) = A_{\tau}p + B_{\tau}, \qquad
    v_{\mathrm{c}}(p)   = A_{v}\exp(B_{v}p).
    \label{eq:TT_params}
\end{equation}
All parameters for the gold-hexadecane system are summarized in Table~\ref{tab:params}.
\begin{table}[htbp]
\centering
\caption{Parameters for the constitutive models from Ref.~\cite{codrignani2023_continuum}}
\label{tab:params}
\begin{tabular}{llcc}
\hline\hline
Model & Parameter & Symbol & Value \\
\hline
\multirow{4}{*}{Tait--Murnaghan, Eq.~\eqref{eq:TaitMurnaghan}}
  & Reference density     & $\rho_{0}$       & $700\ \mathrm{kg\,m^{-3}}$ \\
  & Reference pressure    & $p_\mathrm{TM}$  & $0.101\ \mathrm{MPa}$      \\
  & Bulk modulus          & $K_\mathrm{TM}$  & $0.557\ \mathrm{GPa}$      \\
  & EOS exponent          & $n_\mathrm{TM}$  & $7.33$                     \\[5pt]
\multirow{4}{*}{Roelands, Eq.~\eqref{eq:Roelands}}
  & Ambient viscosity     & $\eta_{0}$       & $0.37\ \mathrm{mPa\,s}$    \\
  & Limiting viscosity    & $\eta_{\infty}$  & $0.06315\ \mathrm{mPa\,s}$ \\
  & Pressure scale        & $p_{\mathrm{R}}$          & $0.1\ \mathrm{GPa}$        \\
  & Roelands exponent     & $z_{\mathrm{R}}$          & $0.51$                     \\[5pt]
\multirow{4}{*}{Carreau, Eq.~\eqref{eq:Carreau}-\eqref{eq:Carreau_params}}
  & Ref.\ shear rate      & $\dot{\gamma}_{00}$ & $1.395\times10^{10}\ \mathrm{s^{-1}}$      \\
  & Pressure scale ($\dot\gamma_{0}$) & $p_{\dot{\gamma}_{0}}$ & $0.5022\ \mathrm{GPa}$  \\
  & Zero-shear exponent   & $n_{0}$          & $0.6216$                                             \\
  & Pressure scale ($n_\mathrm{Car}$) & $p_{n_\mathrm{Car}}$ & $2.028\ \mathrm{GPa}$  \\[5pt]
\multirow{4}{*}{Thompson--Troian, Eq.~\eqref{eq:TT}-\eqref{eq:TT_params}}
  & Stress slope     & $A_{\tau}$  & $8.058\times10^{-3}$ \\
  & Stress intercept & $B_{\tau}$  & $17.44\ \mathrm{MPa}$ \\
  & Velocity prefactor & $A_{v}$   & $24.126\ \mathrm{m\,s^{-1}}$ \\
  & Velocity exponent  & $B_{v}$   & $-2.555\ \mathrm{GPa^{-1}}$ \\[4pt]
\hline\hline
\end{tabular}
\end{table}

\clearpage

\section{Supplementary figures}

\begin{figure}[!ht]
    \centering
    \includegraphics[width=\linewidth]{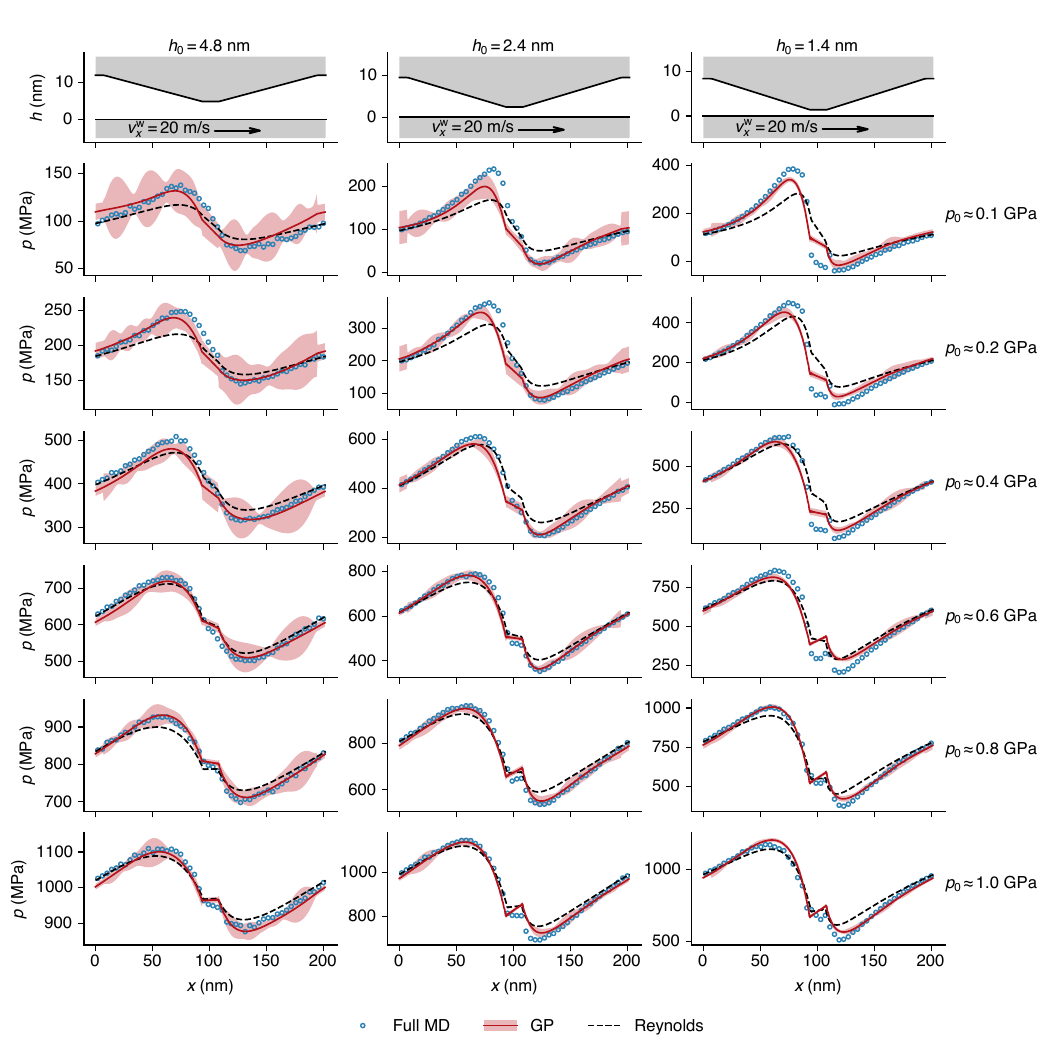}
    \caption{%
        \textbf{Pressure distribution of the converging-diverging channel (CDC) geometry.}
        The first row shows the CDC gap profile for three minimum gap heights $h_0\in[4.8, 2.4, 1.4]\,\mathrm{nm}$ and the columns below show pressure profiles for six normal loads $p_0\in[0.1, 0.2, 0.4, 0.6, 0.8, 1.0]\,\mathrm{GPa}$.
        Molecular dynamics (MD) results from Ref.~\cite{codrignani2023_continuum} are shown as blue circles.
        Pressure distributions obtained by solving the Reynolds equation with the constitutive laws from Ref.~\cite{codrignani2023_continuum} are shown as black dashed lines.
        Solid red lines represent the posterior mean function of the Gaussian process (GP) multiscale simulations, while the shaded red area indicates the 95\% confidence interval given by the GP posterior variance.
    }
    \label{fig:pressure_full}
\end{figure}

\begin{figure}[!ht]
    \centering
    \includegraphics[width=\linewidth]{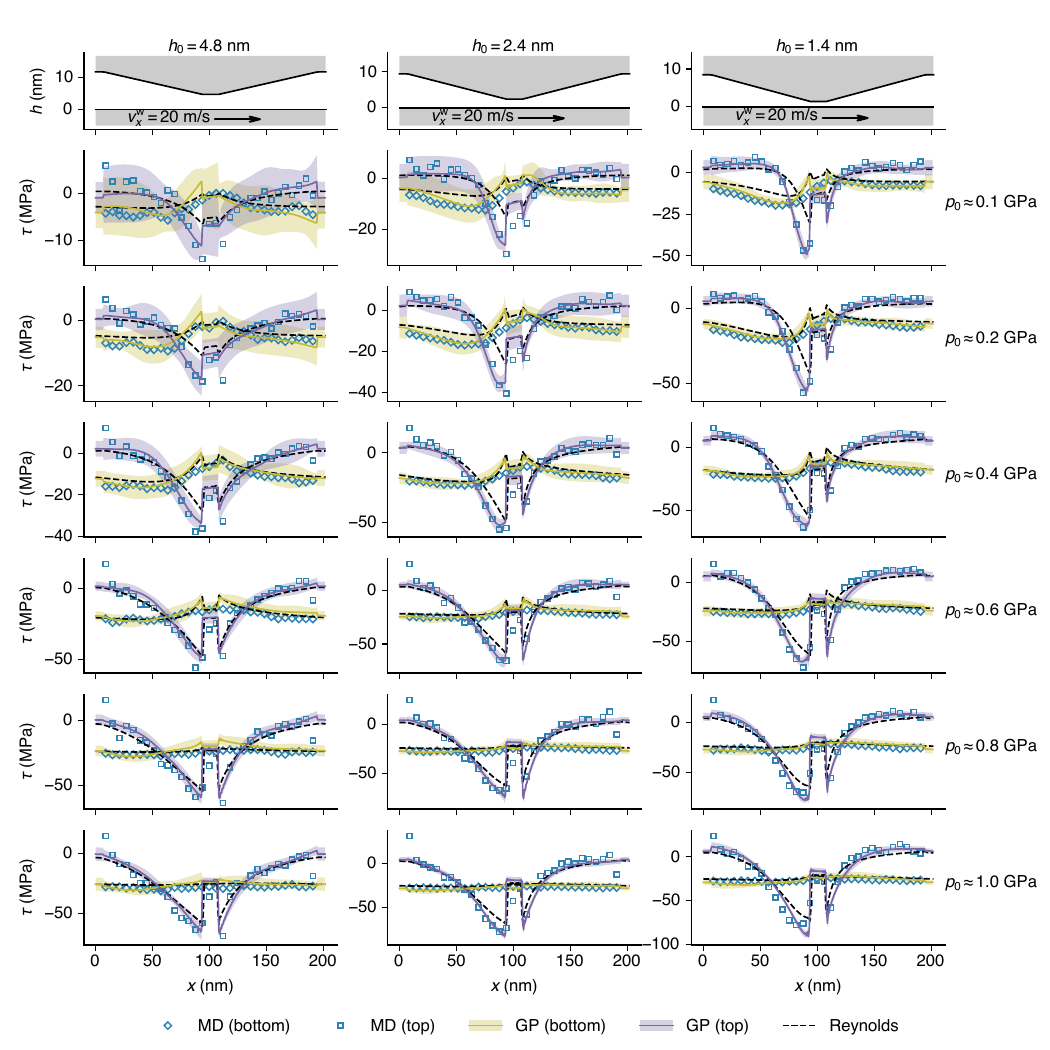}
    \caption{%
        \textbf{Shear stress distribution of the converging-diverging channel (CDC) geometry.}
        The first row shows the CDC gap profile for three minimum gap heights $h_0\in[4.8, 2.4, 1.4]\,\mathrm{nm}$ and the columns below show shear stress profiles at the bottom and top walls for six normal loads $p_0\in[0.1, 0.2, 0.4, 0.6, 0.8, 1.0]\,\mathrm{GPa}$.
        Molecular dynamics (MD) results from Ref.~\cite{codrignani2023_continuum} are shown as blue circles and squares for the bottom and top walls, respectively.
        Shear stress distributions obtained by solving the Reynolds equation with the constitutive laws from Ref.~\cite{codrignani2023_continuum} are shown as black dashed lines.
        Solid red and green lines represent the posterior mean function of the Gaussian process (GP) multiscale simulations for bottom and top walls, respectively.
        The corresponding red and green shaded areas indicate the 95\% confidence interval given by the GP posterior variance.
    }
    \label{fig:shear_full}
\end{figure}

\begin{figure}[!ht]
    \centering
    \includegraphics[width=\linewidth]{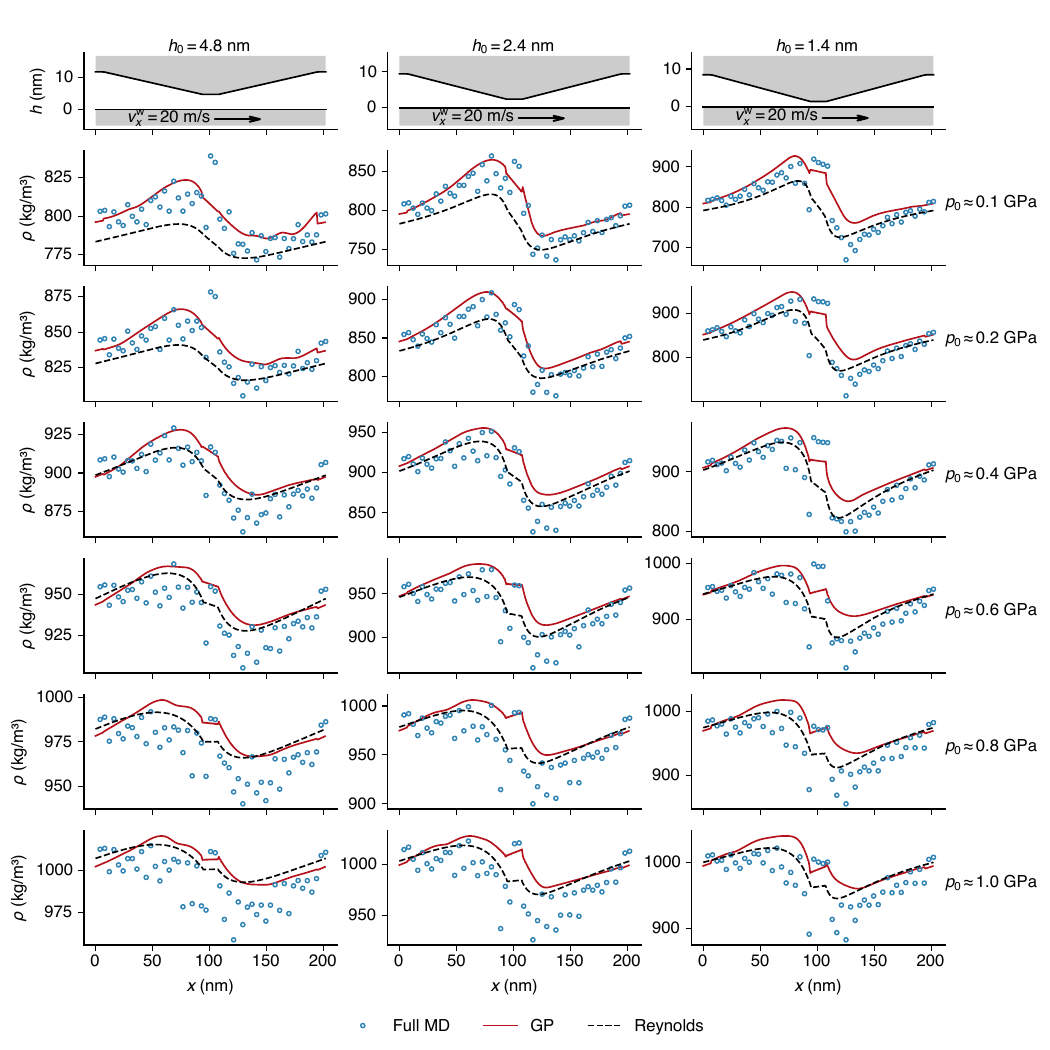}
    \caption{%
        \textbf{Density distribution of the converging-diverging channel (CDC) geometry.}
        The first row shows the CDC gap profile for three minimum gap heights $h_0\in[4.8, 2.4, 1.4]\,\mathrm{nm}$ and the columns below show density profiles for six normal loads $p_0\in[0.1, 0.2, 0.4, 0.6, 0.8, 1.0]\,\mathrm{GPa}$.
        Molecular dynamics (MD) results from Ref.~\cite{codrignani2023_continuum} are shown as blue circles.
        Density distributions obtained by solving the Reynolds equation with the constitutive laws from Ref.~\cite{codrignani2023_continuum} are shown as black dashed lines.
        Solid red lines represent the density distribution obtained with the Gaussian process (GP) multiscale simulations.  
    }
    \label{fig:density_full}
\end{figure}

\begin{figure}[!ht]
    \centering
    \includegraphics[width=\linewidth]{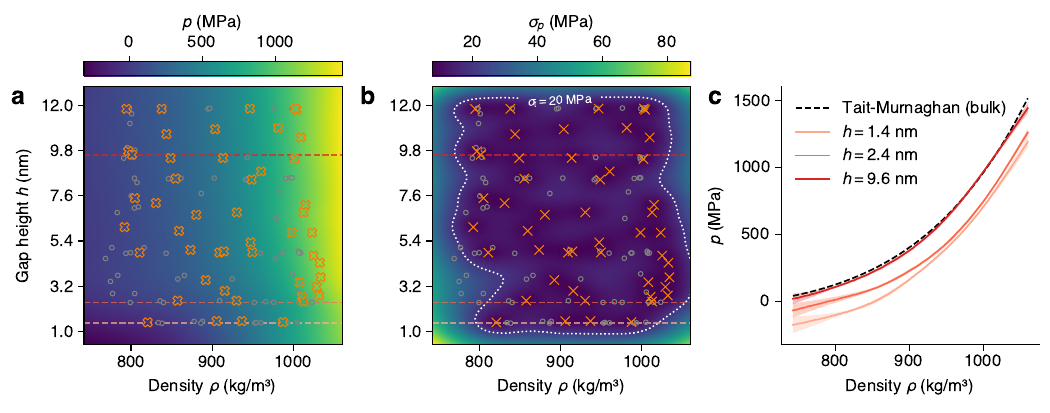}
    \caption{%
        \textbf{Equation of state for heterogeneous, converging channels.}
        \textbf{a} Two-dimensional slice through the three-dimensional input space at $\dd h/\dd x<0$, i.e. for heterogeneous channels in the converging part of the CDC. Training data at these conditions is highlighted as open orange crosses. The color filling of the symbols corresponds to the measured MD pressure. Training data from flat and diverging channels are shown as gray circles.
        \textbf{b} Uncertainty of the pressure model illustrated as the standard deviation $\sigma_p$. Orange and gray symbols correspond to the same training data as in \textbf{a}. The white dotted line highlights the boundary between ``accepted'' and ``non-accepted'' regions in an exemplary active learning simulation with uncertainty tolerance $\sigma_\mathrm{t}=20\,\mathrm{MPa}$. 
        \textbf{c} Equation of state at constant gap heights illustrated by the horizontal dashed lines in \textbf{a} and \textbf{b}. The black dashed line highlights the Tait-Murnaghan equation of state as parametrized in Ref.~\cite{codrignani2023_continuum} with bulk MD simulations.
    }
    \label{fig:supp_eos_converging}
\end{figure}

\end{document}